\documentclass[useAMS,usenatbib]{mn2e}
\pdfoutput=1

\newcommand{\wrot}{\omega_{\text{rot}}}
\newcommand{\hwrot}{\hat{\omega}_{\text{rot}}}
\newcommand{\worb}{\omega_{\text{orb}}}
\newcommand{\hworb}{\hat{\omega}_{\text{orb}}}
\newcommand{\thobs}{\theta_{o}}
\newcommand{\phobs}{\phi_{o}}
\newcommand{\thst}{\theta_{s}}
\newcommand{\phst}{\phi_{s}}
\newcommand{\xis}{\xi_{\text{s}}}
\newcommand{\phobsz}{\phi_{o}(0)}
\newcommand{\rps}{\hat{r}_{\text{ps}}}
\newcommand{\kwidth}{\sigma_{\phi}(\mathbb{G})}
\newcommand{\kdomlat}{\bar{\theta}(\mathbb{G})}
\newcommand{\ktwod}{K(\theta,\phi,\mathbb{G})}
\newcommand{\thspec}{\theta_{\text{spec}}}
\newcommand{\phispec}{\phi_{\text{spec}}}

\usepackage{times}
\usepackage{graphicx}
\usepackage{xcolor}
\usepackage{amssymb}
\usepackage{amsmath}

\title[Inferring Planetary Obliquity]{Inferring Planetary Obliquity Using Rotational \& Orbital Photometry}
\author[J. C. Schwartz et al.]{J. C. Schwartz$^{1,2,3}$\thanks{E-mail: joelschwartz2011@u.northwestern.edu}\thanks{McGill Space Institute (McGill U.); Institute for Research on Exoplanets (UdeM)}, C. Sekowski$^{4,5}$, H. M. Haggard$^{4}$, E. Pall{\'e}$^{6}$, and N. B. Cowan$^{2,3}$$\dagger$\\
$^{1}$Department of Physics \& Astronomy, Northwestern University, 2145 Sheridan Road, Evanston, IL, 60208, USA\\
$^{2}$Department of Earth \& Planetary Sciences, McGill University, 3450 rue University, Montreal, QC, H3A 0E8, CAN\\
$^{3}$Department of Physics, McGill University, 3600 rue University, Montreal, QC, H3A 2T8, CAN\\
$^{4}$Physics Program, Bard College, PO Box 5000, Annandale, NY, 12504, USA\\
$^{5}$Department of Physics, Boston University, 590 Commonwealth Ave, Boston, MA, 02215, USA\\
$^{6}$Instituto de Astrof{\'i}scia de Canarias, Via L{\'a}ctea s/n, La Laguna, Santa Cruz de Tenerife, 35205, ESP}
\begin{document}

\date{Published in MNRAS}

\pagerange{\pageref{firstpage}--\pageref{lastpage}}\pubyear{2015}

\maketitle

\label{firstpage}

\begin{abstract}
The obliquity of a terrestrial planet is an important clue about its formation and critical to its climate. Previous studies using simulated photometry of Earth show that continuous observations over most of a planet's orbit can be inverted to infer obliquity. However, few studies of more general planets with arbitrary albedo markings have been made and, in particular, a simple theoretical understanding of why it is possible to extract obliquity from light curves is missing. Reflected light seen by a distant observer is the product of a planet's albedo map, its host star's illumination, and the visibility of different regions. It is useful to treat the product of illumination and visibility as the kernel of a convolution. Time-resolved photometry constrains both the albedo map and the kernel, the latter of which sweeps over the planet due to rotational and orbital motion. The kernel's movement distinguishes prograde from retrograde rotation for planets with non-zero obliquity on inclined orbits. We demonstrate that the kernel's longitudinal width and mean latitude are distinct functions of obliquity and axial orientation.  Notably, we find that a planet's spin axis affects the kernel---and hence time-resolved photometry---even if this planet is East-West uniform or spinning rapidly, or if it is North-South uniform. We find that perfect knowledge of the kernel at 2--4 orbital phases is usually sufficient to uniquely determine a planet's spin axis. Surprisingly, we predict that East-West albedo contrast is more useful for constraining obliquity than North-South contrast.
\end{abstract}


\section{Introduction}
\label{sec:intro}
The obliquity of a terrestrial planet encodes information about different processes. A planet's axial alignment and spin rate inform its formation scenario. Numerical simulations have shown that the spin rates of Earth and Mars are likely caused by a few planetesimal impacts \citep{dones1993origin}, while perfect accretion produces an obliquity distribution that is isotropic \citep[e.g.][]{kokubo2007formation,miguel2010planet}. Conversely, \cite{schlichting2007effect} describe how prograde rotation is preferred to retrograde for a formation model with semi-collisional accretion.

Obliquity is also important in controlling planetary climate. This has been studied in-depth for Earth under many conditions \citep[e.g.][]{laskar2004long,pierrehumbert2010principles}, and high axial tilts can make planets at large semi-major axes more habitable \citep{williams1997habitable}. Furthermore, while the Earth's spin axis is stabilized by the Moon \citep{laskar1993stabilization}, obliquities of several Solar System bodies evolve chaotically \citep{laskar1994large}. This influences searches for hospitable planets, as \cite{spiegel2009habitable} note that the habitability of terrestrial worlds may depend sensitively on how stable the climate is in the short-term.

\begin{figure*}
	\centering
	\includegraphics[width=1.0\linewidth]{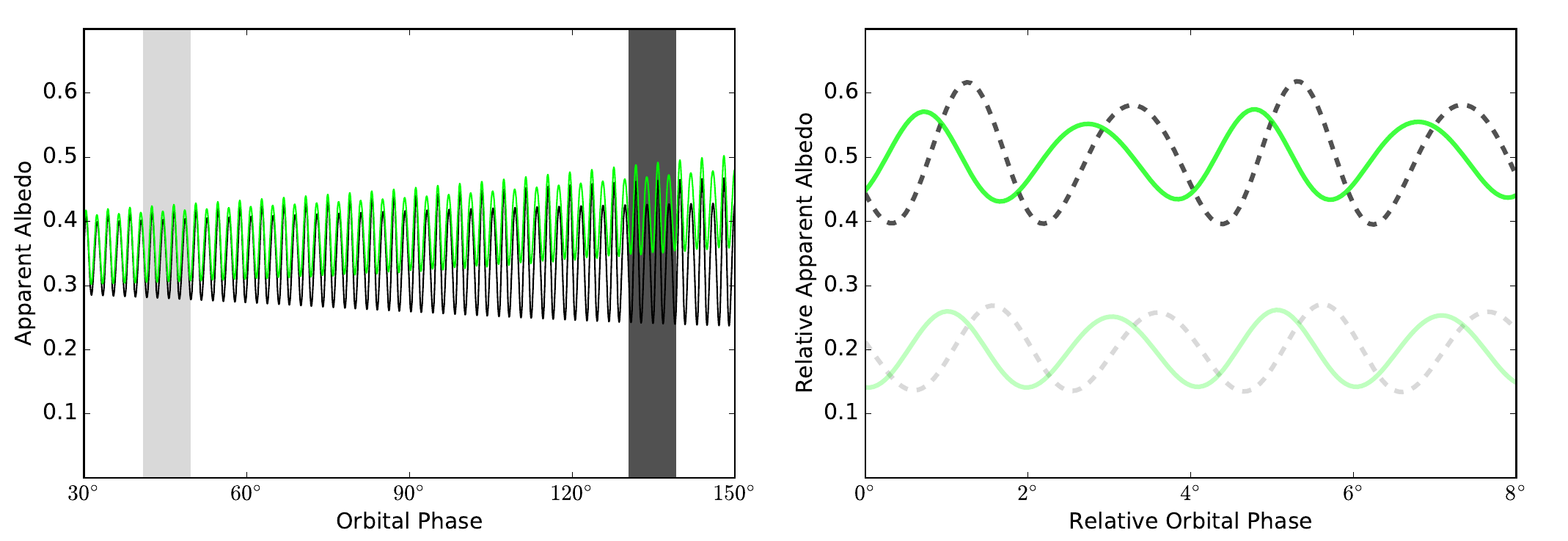}
	\caption{The left panel shows apparent albedo as a function of orbital phase for an arbitrary planet with North-South and East-West albedo markings, seen edge-on with zero obliquity (black) or $45^{\circ}$ obliquity (green). The average albedo of the green planet increases during the orbit because brighter latitudes become visible and illuminated; this does not happen for the black planet. The vertical bands are each roughly two planetary days, enlarged at right, where lighter shades are the fuller phase. For clarity, the rotational curves are shifted and the zero-obliquity planet is denoted by a dashed line. The apparent albedo of both planets varies more over a day when a narrower range of longitudes and albedo markings are visible and illuminated, and vice versa. Note that both light curves are produced by the \emph{same} albedo map, and are distinct solely because of differences in the kernel for these two geometries. Orbital and/or rotational changes in apparent albedo can help one infer a planet's obliquity.}
	\label{fig:ex_LCs_edge_0q45q}
\end{figure*}

A planet's average insolation is set by stellar luminosity and semi-major axis; insolation at different latitudes is determined by obliquity and (for eccentric orbits) the axial orientation. Non-oblique planets have a warmer equator and colder poles that do not vary much throughout the year. Modest obliquities produce seasons at mid-latitudes because the sub-stellar point moves North and South during the orbit \citep{pierrehumbert2010principles}. Planets tilted at \mbox{angles $\geq 54^{\circ}$} receive more overall radiation near their poles and have large orbital variations in temperature \citep{williams2003extraordinary}. Thus, even limited knowledge of a planet's obliquity can help constrain the spatial dependence of insolation and temperature.

Numerous methods have been proposed for measuring planetary obliquities. \citet{seager2002constraining} and \citet{barnes2003measuring} demonstrated constraints on oblateness and obliquity using ingress/egress differences in transit light curves; \citet{carter2010empirical} extended and applied these techniques to observations of HD 189733b. \citet{kawahara2012spin} derived constraints on obliquity from modulation of a planet's radial velocity during orbit, while \citet{nikolov2015radial} examined the Rossiter-McLauglin effect at secondary eclipse for transiting exoplanets. One could also measure obliquity at infrared wavelengths, using polarized rotational light curves \citep{de2011characterizing} and orbital variations \citep[e.g.][]{gaidos2004seasonality,cowan2013light}.

A planet's obliquity can also be constrained by changes in reflected light, which will be studied with forthcoming optical and near-infrared space missions, such as \emph{\mbox{ATLAST}} \citep{postman2010advanced}, \emph{\mbox{LUVOIR}} \citep{kouveliotou2014enduring}, and \emph{\mbox{HDST}} \citep{dalcanton2015cosmic}. Time-resolved measurements of a rotating planet in one photometric band can reveal its rotation rate \citep{ford2001characterization,palle2008identifying,oakley2009construction}; this helps determine Coriolis forces and predict large-scale circulation. Multi-band photometry can reveal colors of clouds and surface features \citep{ford2001characterization,fujii2010colors,fujii2011colors,cowan2013determining}, and enables a longitudinal albedo map to be inferred from disk-integrated light \citep{cowan2009alien}. High-cadence, reflected light measurements spanning a full planetary orbit constrain a planet's obliquity and two-dimensional albedo map \citep{kawahara2010global,kawahara2011mapping,fujii2012mapping}.

However, these results have not yet been established for lower cadence measurements of non-terrestrial planets. We also hope to establish a conceptual understanding of how photometric measurements constrain obliquity. While this is less immediately practical, a deeper understanding of this inversion has the potential to lead to further advances in inferring planetary geometry from limited data sets. In this paper, we study light curve methods for arbitrary albedo maps and viewing geometries, and demonstrate they are useful even for observations at only one or two orbital phases.

Light curves of planets encode the viewing geometry and hence a planet's obliquity because different latitudes are impinged by starlight at different orbital phases (this ``kernel" is described in Section \ref{sec:kern}). To see this, consider a planet with no obliquity in an edge-on, circular orbit. The star always illuminates the Northern and Southern hemispheres equally, and we never view some latitudes more than others. If instead this planet were tilted, the Northern hemisphere would be lit first, then the Southern hemisphere half an orbit later. If the planet is not North-South uniform, its apparent albedo \citep{qui2003earthshine,cowan2009alien} would change during its orbit, shown in the left panel of Figure \ref{fig:ex_LCs_edge_0q45q}.

One may also learn about a planet's obliquity as it rotates. Imagine a zero-obliquity planet in a face-on, circular orbit: the observer always sees the Northern pole with half the longitudes illuminated. For an oblique planet, however, more longitudes would be lit when the visible pole leans towards the star, and vice versa. Zero-obliquity planets in edge-on orbits are similar, since more longitudes are lit near superior conjunction, or fullest phase. If the planet has East-West albedo variations, then this longitudinal width will modulate the apparent albedo of the planet as it spins, shown in the right panel of Figure \ref{fig:ex_LCs_edge_0q45q}.

Our work is organized as follows: in Section \ref{sec:reflig}, we summarize the observer viewing geometry and explain the reflective kernel, both in two- and one-dimensional forms. Section \ref{sec:ind_pha} introduces a case study planet and describes the kernel at single orbital phases; we consider time evolution in Section \ref{sec:t_evo}. We discuss real observations in Section \ref{sec:observing}, then develop our case study in \mbox{Sections \ref{sec:long_con}--\ref{sec:joint_con}}, demonstrating that even single- and dual-epoch observations could allow one to constrain obliquity. In Section \ref{sec:pro_ret}, we discuss how to distinguish a planet's rotational direction by monitoring its apparent albedo. Section \ref{sec:conclude} summarizes our conclusions. For interested readers, a full mathematical description of the illumination and viewing geometry is presented in \mbox{Appendix \ref{sec:v_geo}}. Details about the kernel and its relation to a planet's apparent albedo are described in Appendix \ref{sec:K_detail}.

\section{Reflected Light}
\label{sec:reflig}

\subsection{Geometry \& Flux}
\label{sec:geom}
The locations on a planet that contribute to the disk-integrated reflected light depend only on the sub-observer and sub-stellar positions, which both vary in time. A complete development of this viewing geometry is provided in Appendix \ref{sec:v_geo}, which we summarize here. We neglect axial precession and consider planets on circular orbits. Assuming a static albedo map, the reflected light seen by an observer is determined by the colatitude and longitude of the sub-stellar and sub-observer points, explicitly $\thst$, $\phst$, $\thobs$, and $\phobs$. The intrinsic parameters of the system are the orbital and rotational angular frequencies, $\worb$ and $\wrot$ (where positive $\wrot$ is prograde), and the planetary obliquity, $\Theta \in [0,\pi/2]$. Extrinsic parameters differ from one observer to the next; we denote the orbital inclination, $i$ (where $i=90^{\circ}$ is edge-on), and solstice phase, $\xis$ (the orbital phase of Summer solstice for the Northern hemisphere). We also define initial conditions for orbital phase, $\xi_{0}$, and the sub-observer longitude, $\phobsz$. Reflected light is then completely specified by these seven parameters and the planet's albedo map.

We consider only diffuse (Lambertian) reflection in our analysis. Specular reflection, or glint, can be useful for detecting oceans \citep{williams2008detecting,robinson2010detecting,robinson2014detection}, but is a localized feature and a minor fraction of the reflected light at gibbous phases. The reflected flux measured by a distant observer is therefore a convolution of the two-dimensional kernel \cite[or weight function;][]{fujii2012mapping}, $K(\theta,\phi,\mathbb{S})$, and the planet's albedo map, $A(\theta,\phi)$:
\begin{equation}
	\label{eq:flux}
	F(t) = \oint K(\theta,\phi,\mathbb{S}) A(\theta,\phi) \text{d}\Omega,
\end{equation}
where $F$ is the observed flux, $\theta$ and $\phi$ are colatitude and longitude, and $\mathbb{S}~\equiv~\lbrace\thst,\phst,\thobs,\phobs\rbrace$ implicitly contains the time-dependencies in the sub-stellar and sub-observer locations. Reconstructing a map of an exoplanet based on time-resolved photometry can be thought of as a deconvolution \citep{cowan2013light}, while estimating a planet's obliquity amounts to backing out the kernel of the convolution.

The sub-stellar and sub-observer points are completely determined through a function $\mathbb{S} = f(\mathbb{G},\wrot t)$, where $\mathbb{G}~\equiv~\left\lbrace\xi(t),i,\Theta,\xis\right\rbrace$ is the system geometry and $\xi(t)$ is orbital phase. This is made explicit in Appendix \ref{sec:v_geo}. For a planet with albedo markings, one would therefore fit the observed flux to infer both the planet's albedo map \citep{cowan2009alien} and spin axis \citep{kawahara2010global,kawahara2011mapping,fujii2012mapping}. To study how these methods work for arbitrary maps and geometries, we will focus on the kernel from Equation \ref{eq:flux}, which we can analyze independent of a planet's albedo map.

\subsection{Kernel}
\label{sec:kern}
The kernel combines illumination and visibility, defined for diffuse reflection in \citet{cowan2013light} as
\begin{equation}
	\label{eq:kern}
	K(\theta,\phi,\mathbb{S}) = \frac{1}{\pi} V(\theta,\phi,\thobs,\phobs) I(\theta,\phi,\thst,\phst),
\end{equation}
where $V(\theta,\phi,\thobs,\phobs)$ is the visibility and $I(\theta,\phi,\thst,\phst)$ is the illumination. Visibility and illumination are each non-zero over one hemisphere at any time, and are further given by
\begin{equation}
	\label{eq:vis}
	\begin{aligned}
	V(\theta,\phi,\thobs,\phobs) = \text{max}\big[&\sin\theta\sin\thobs\cos(\phi - \phobs)\\
	&+ \cos\theta\cos\thobs, 0\big],
	\end{aligned}
\end{equation}
\begin{equation}
	\label{eq:illum}
	\begin{aligned}
	I(\theta,\phi,\thst,\phst) = \text{max}\big[&\sin\theta\sin\thst\cos(\phi - \phst)\\
	&+ \cos\theta\cos\thst, 0\big].
	\end{aligned}
\end{equation}
As noted above, we can express the kernel equivalently as
\begin{equation}
	\label{eq:kern_G}
	K(\theta,\phi,\mathbb{S}) = K(\theta,\phi,\mathbb{G},\wrot t),
\end{equation}
though we will drop the rotational dependence for now because it does not affect our analysis. We return to rotational frequency in Section \ref{sec:pro_ret}.

\begin{figure}
	\centering
	\includegraphics[width=1.0\linewidth]{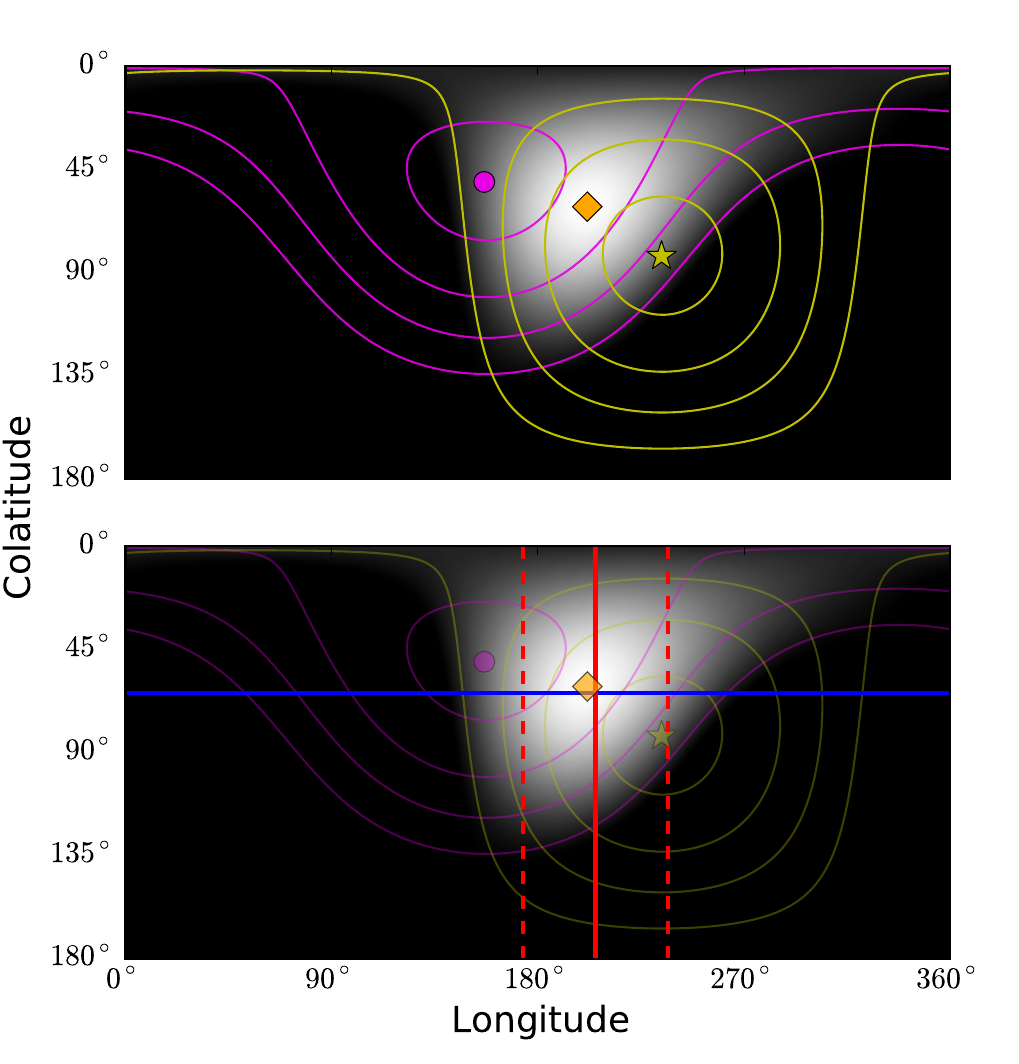}
	\caption{\emph{Upper:} A kernel, in gray, with contours showing visibility and illumination, in purple and yellow, as in \citet{cowan2013light}. The sub-observer and sub-stellar points are indicated by the purple circle and yellow star, respectively. The orange diamond marks the peak of the kernel. \emph{Lower:} The mean of the longitudinal kernel and the width from this mean are shown as solid and dashed red lines; the dominant colatitude is shown as a blue line.}
	\label{fig:ex_kern_0xis_2panel}
\end{figure}

The non-zero portion of the kernel is a lune: the illuminated region of the planet that is visible to a given observer. The size of this lune depends on orbital phase, or the angle between the sub-observer and sub-stellar points. A sample kernel is shown at the top of Figure \ref{fig:ex_kern_0xis_2panel}, where the purple and yellow contours are visibility and illumination, respectively. The peak of the kernel is marked with an orange diamond.

We begin by calculating time-dependent sines and cosines of the sub-observer and sub-stellar angles for a viewing geometry of interest (Appendix \ref{sec:v_geo}). These are substituted into \mbox{Equations \ref{eq:vis} and \ref{eq:illum}} to determine visibility and illumination at any orbital phase. The two-dimensional kernel is then calculated on a $101 \times 201$ grid in colatitude and longitude.

\subsection{Longitudinal Width}
\label{sec:LW}

\begin{figure*}
	\centering
	\includegraphics[width=1.0\linewidth]{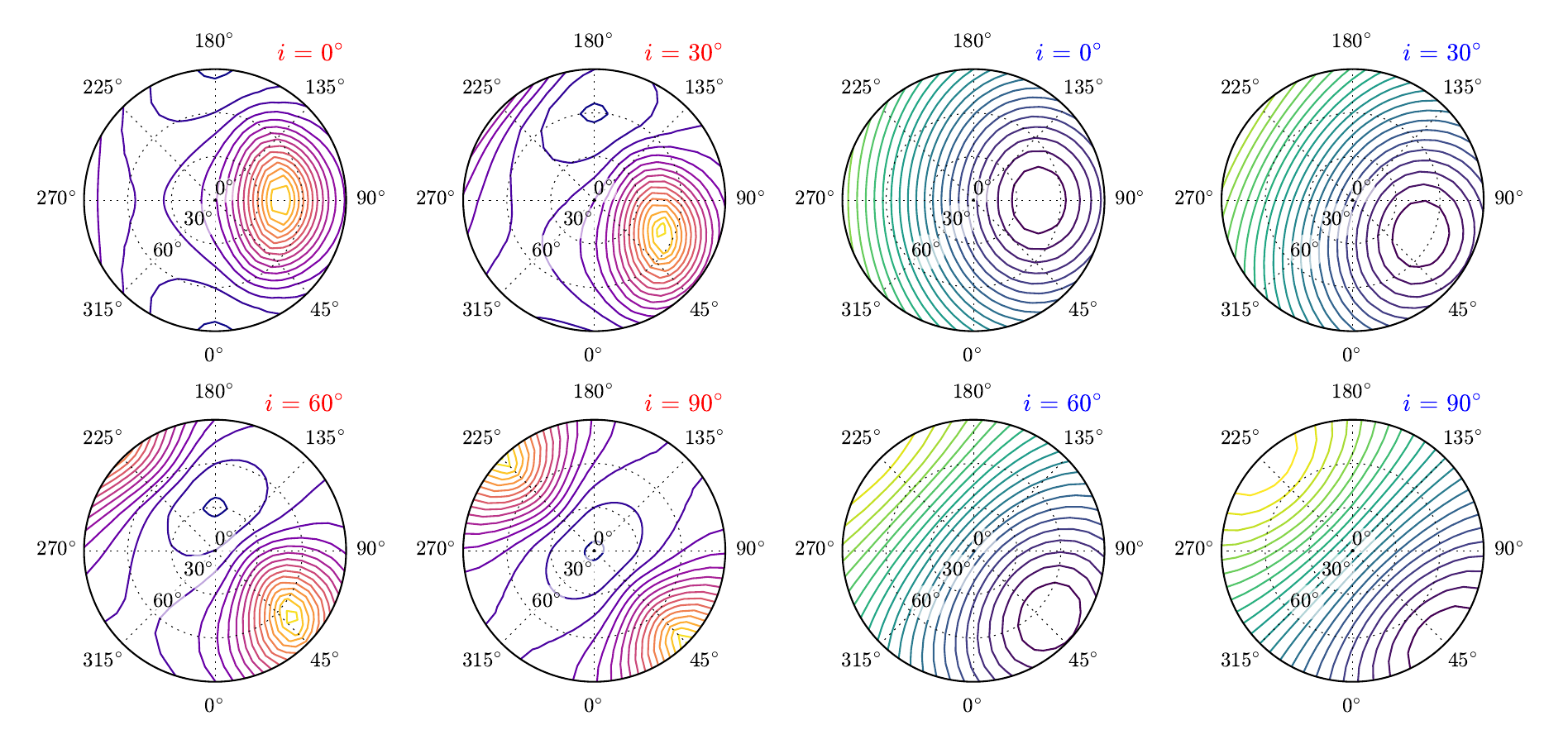}
	\caption{\emph{Left panels:} Contours of longitudinal kernel width at first quarter phase, $\xi(t) = 90^{\circ}$, as a function of planetary obliquity, from face-on $i=0^{\circ}$ to edge-on $i=90^{\circ}$. Obliquity is plotted radially: the center is $\Theta = 0^{\circ}$ and the edge is $\Theta = 90^{\circ}$. The azimuthal angle represents the orientation (solstice phase) of the planet's obliquity. The contours span $20^{\circ}$--$100^{\circ}$, dark to light colors, in $5^{\circ}$ increments. Larger kernel widths---and hence muted rotational variability---occur at gibbous phases, and when the peak of the kernel is near a pole. \emph{Right panels:} Analogous contours of dominant colatitude that \mbox{span $35^{\circ}$--$145^{\circ}$}. A lower dominant colatitude---and thus more reflection from the Northern hemisphere---occurs when the kernel samples Northern regions more, and vice versa.}
	\label{fig:ex_Kphith_contour_oct}
\end{figure*}

The two-dimensional kernel, $\ktwod$, is a function of latitude and longitude that varies with time and viewing geometry. For observations with minimal orbital coverage or planets that are North-South uniform, different latitudes are hard to distinguish \citep{cowan2013light} and we use the longitudinal form of the \mbox{kernel, $K(\phi, \mathbb{G})$,} given by
\begin{equation}
	\label{eq:K_phi}
	K(\phi,\mathbb{G}) = \int_{0}^{\pi} K(\theta,\phi,\mathbb{G})\sin\theta\text{d}\theta.
\end{equation}
We can approximately describe $K(\phi, \mathbb{G})$ by a longitudinal mean, $\bar{\phi}$, and width, $\sigma_{\phi}$. These are defined in Appendix \ref{sec:charact}; examples are shown as vertical red lines in the bottom panel of Figure \ref{fig:ex_kern_0xis_2panel}.

For any geometry, we can calculate the two-dimensional kernel and the corresponding longitudinal width. The mean longitude is unimportant by itself because, for now, we are only concerned with the size of the kernel. We compute a four-dimensional grid of kernel widths with $5^{\circ}$ resolution in orbital phase (time), inclination, obliquity, and solstice phase. The result is ${\sigma_{\phi}(\xi(t),i,\Theta,\xis) \equiv \kwidth}$, and our numerical grid has size ${73 \times 19 \times 19 \times 73}$ in the respective parameters. Example contours from this array at first quarter phase, or $\xi(t)=90^{\circ}$, are shown in the left panels of Figure \ref{fig:ex_Kphith_contour_oct}. In these plots obliquity is radial: the center is $\Theta = 0^{\circ}$ and the edge is $\Theta = 90^{\circ}$. The azimuthal angle gives the orientation (solstice phase) of the planet's obliquity.

\subsection{Dominant Colatitude}
\label{sec:DC}
For a given planet and observer, the sub-observer colatitude is fixed but the sub-stellar point moves North and South throughout the orbit if the planet has non-zero obliquity. This means different orbital phases will probe different latitudes, as dictated by the kernel. To analyze these variations we use the latitudinal form of the kernel, $K(\theta,\mathbb{G})$, explicitly:
\begin{equation}
	\label{eq:K_theta}
	K(\theta,\mathbb{G}) = \int_{0}^{2\pi} K(\theta,\phi,\mathbb{G})\text{d}\phi.
\end{equation}
We may describe $K(\theta,\mathbb{G})$ by its dominant colatitude \citep{cowan2012false}, $\bar{\theta}$, also defined in Appendix \ref{sec:charact} and shown as a horizontal blue line at the bottom of Figure \ref{fig:ex_kern_0xis_2panel}. We produce a four-dimensional dominant colatitude array, $\bar{\theta}(\xi(t),i,\Theta,\xis)~\equiv~\kdomlat$, similarly to $\kwidth$ from Section \ref{sec:LW}. Sample contours from this array at first quarter phase are shown in the right panels of Figure \ref{fig:ex_Kphith_contour_oct}.

\section{Kernel Behavior}
\label{sec:K_beh}
We now consider how the longitudinal width and dominant colatitude of the kernel depend on a planet's obliquity. As a case study, we will define the inclination and spin axis of a hypothetical planet, $Q$:
\begin{equation}
	\label{eq:pl_Q}
	\text{planet} \ Q \equiv
	\left\lbrace
		\begin{aligned}
		i &= 60^{\circ}\\ 
		\Theta &= 55^{\circ}\\
		\xis &= 260^{\circ}
	\end{aligned}
	\right\rbrace.
\end{equation}

\subsection{Phases}
\label{sec:ind_pha}

\begin{figure*}
	\centering
	\includegraphics[width=1.0\linewidth]{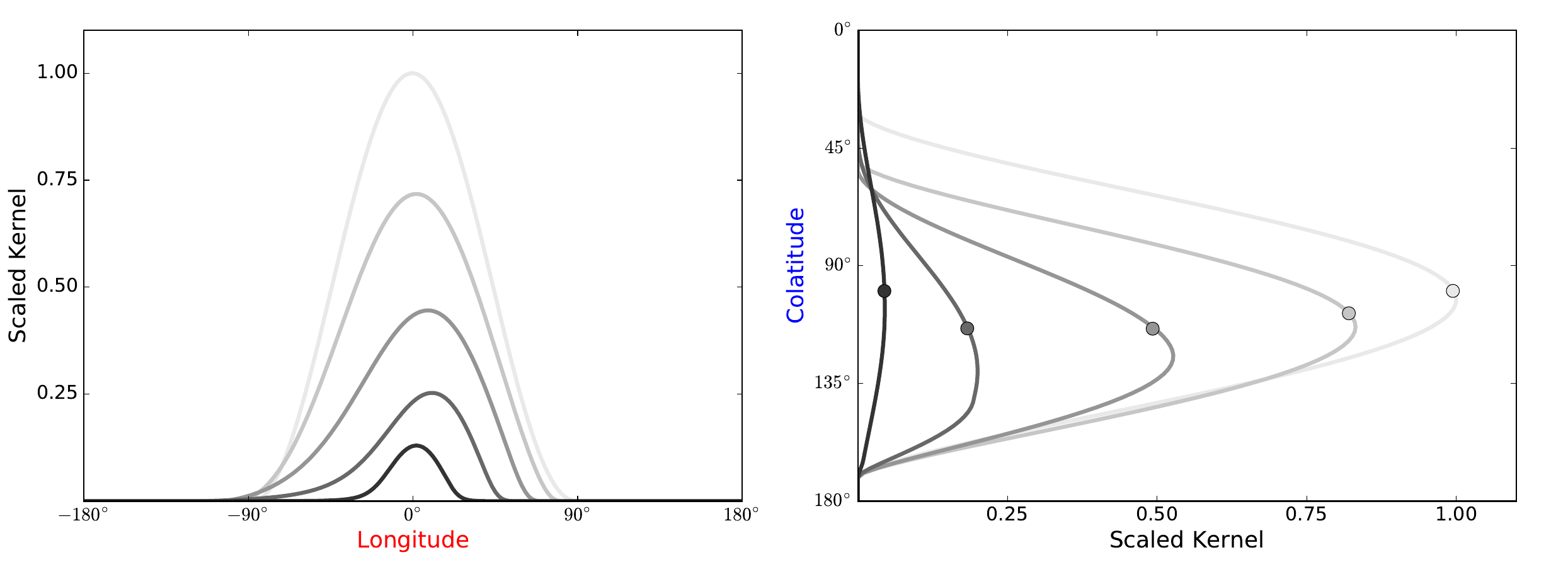}
	\caption{\emph{Left:} Longitudinal kernel for planet $Q$, defined in Equation \ref{eq:pl_Q}, at orbital phases from $30^{\circ}$ to $150^{\circ}$, light to dark shades, in $30^{\circ}$ increments. Values are scaled to the maximum of the lightest curve. Longitude is measured from each kernel mean. The kernel width decreases as this planet approaches inferior conjunction, or as color darkens. \emph{Right:} Analogous latitudinal kernel for planet $Q$, where the dominant colatitude, indicated by a circle, increases then returns towards the equator.}
	\label{fig:ex_Kphith_55_260_phases30}
\end{figure*}

Considering a single orbital phase defines a three-dimensional slice through $\kwidth$ and $\kdomlat$ that describes the kernel at that specific time. We show the longitudinal form of the kernel for planet $Q$ at different phases in the left panel of Figure \ref{fig:ex_Kphith_55_260_phases30}. Lighter colors are fuller phases, indicating the kernel narrows as this planet orbits towards inferior conjunction, or $\xi(t)=180^{\circ}$. The kernel width influences the rotational light curve at a given phase: narrower kernels can have larger amplitude variability in apparent albedo on a shorter timescale (e.g. right of Figure \ref{fig:ex_LCs_edge_0q45q}).

The latitudinal kernel for planet $Q$ is shown similarly in the right panel of Figure \ref{fig:ex_Kphith_55_260_phases30}. We see that the kernel preferentially probes low and mid-latitudes during the first half-orbit. The dominant colatitude of planet $Q$, indicated by circles, also fluctuates during this portion of the orbit---and eventually shifts well into the Northern hemisphere after inferior conjunction (not shown). Note that the dominant colatitude is \emph{not} always at the peak of the latitudinal kernel (see also Figure \ref{fig:ex_Kphith_225sol_obq}). Since one needs measurements at multiple phases to be sensitive to latitudinal variations in albedo, we will consider \emph{changes} in dominant colatitude from one phase to the next, $|\Delta\bar{\theta}|$, for planets with North-South albedo markings. Larger changes in dominant colatitude can make the apparent albedo vary more between orbital phases (e.g. left of Figure \ref{fig:ex_LCs_edge_0q45q}).

\subsection{Time Evolution}
\label{sec:t_evo}

\begin{figure*}
	\centering
	\includegraphics[width=1.0\linewidth]{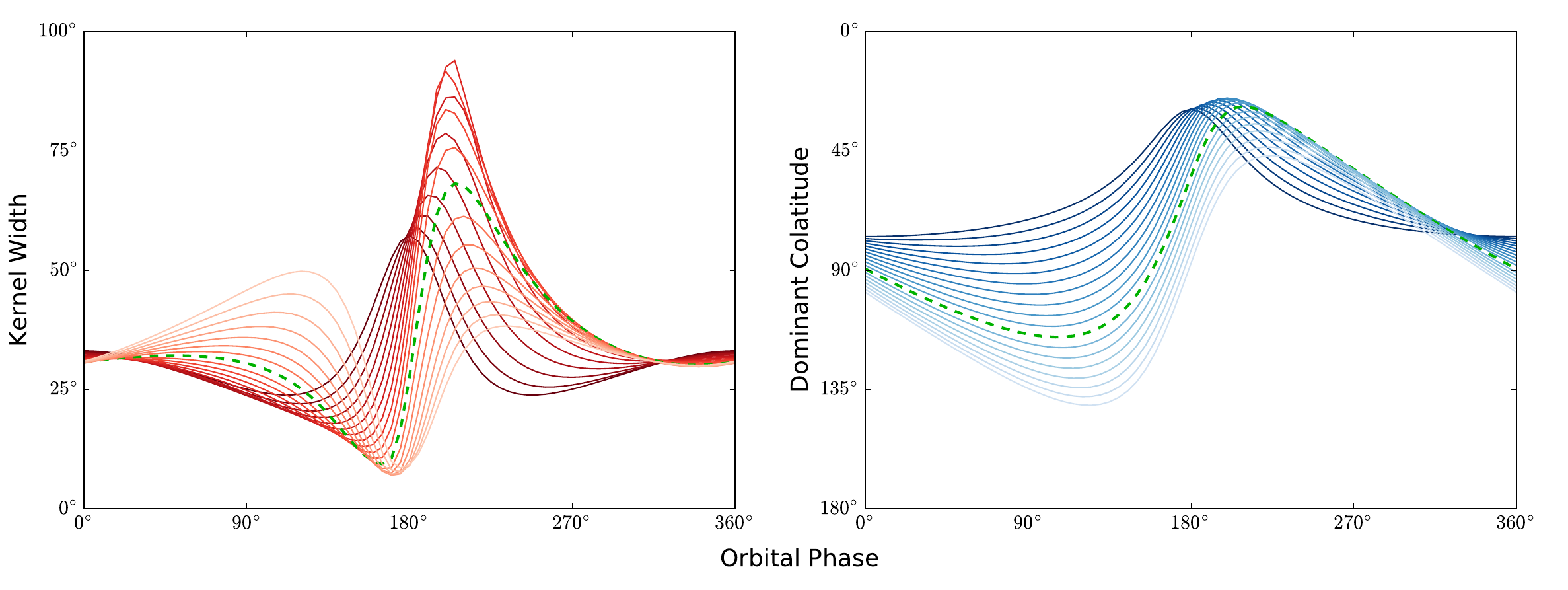}
	\caption{\emph{Left:} Kernel width for planet $Q$ as a function of orbital phase, with obliquity varied in $5^{\circ}$ increments. The defined planet $Q$ obliquity, $\Theta = 55^{\circ}$, is the dashed green line; darker and lighter shades of red denote obliquities closer to $0^{\circ}$ and $90^{\circ}$, respectively. Inferior conjunction occurs at $\xi(t) = 180^{\circ}$. The largest variations are near $\xi(t) \approx \lbrace 120^{\circ},210^{\circ} \rbrace$. \emph{Right:} Analogous dominant colatitude for planet $Q$. Dual-epoch changes are diverse between $\xi(t) \approx \lbrace 135^{\circ},240^{\circ} \rbrace$, for example.}
	\label{fig:ex_qtrace_wl260ph}
\end{figure*}

Kernel width and dominant colatitude both vary throughout a planet's orbit. We investigate this by slicing $\kwidth$ and $\kdomlat$ along obliquity and/or solstice phase. To start, we vary planet $Q$'s obliquity and track kernel width as shown in the left panel of Figure \ref{fig:ex_qtrace_wl260ph}. The actual planet $Q$ is denoted by a dashed green line: this planet has a narrow kernel width during the first half-orbit that widens sharply after inferior conjunction. The largest variations between the traces occur near $\xi(t) \approx \lbrace 120^{\circ},210^{\circ} \rbrace$.

We also show tracks of dominant colatitude in the right panel of Figure \ref{fig:ex_qtrace_wl260ph}. What matters is the change in this characteristic between two epochs; diverse changes in the traces occur between $\xi(t) \approx \lbrace 135^{\circ},240^{\circ} \rbrace$. Planet $Q$ is again the dashed green line, and near the middle of all the tracks more often than for kernel width. If one has some prior knowledge of the viewing geometry, then \mbox{Figure \ref{fig:ex_qtrace_wl260ph}} implies at which phases one could observe to best distinguish obliquities for planet $Q$---for example, $\xi(t) \approx \lbrace 120^{\circ},240^{\circ} \rbrace$.

We can instead vary the solstice phase of planet $Q$ while keeping its obliquity fixed (not shown). In most cases, solstice phase impacts the kernel width and dominant colatitude as much as the axial tilt. This is expected, since obliquity is a vector quantity with both magnitude and orientation.

\section{Discussion}
\label{sec:discuss}

\subsection{Observations}
\label{sec:observing}
By analyzing the kernel, we can learn how observed flux may depend on a planet's obliquity independent from its albedo map. For real observations, one would fit the light curve to directly infer the planet's albedo map \citep{cowan2009alien} and spin axis \citep{kawahara2010global,kawahara2011mapping,fujii2012mapping}. We will use the kernel to \emph{predict} how single- and dual-epoch observations constrain planetary obliquity. We address our assumptions below.

The planetary inclination and orbital phase of observation must be known to model the light curve accurately. Both angles might be obtained with a mixture of astrometry on the host star \citep[e.g. \emph{SIM PlanetQuest};][]{unwin2008taking}, direct-imaging astrometry \citep{bryden2015tracing}, and/or radial velocity. We will assume that inclination and orbital phase have each been measured with  $10^{\circ}$ uncertainty.

Extracting the albedo map and spin axis from a light curve could also be difficult in practice. Planets with completely uniform albedo are not amenable to these methods. Moreover, one cannot distinguish latitudes for planets that are North-South uniform, nor longitudes for those that are East-West uniform. Even if a planet has albedo contrast, photometric uncertainty adds noise to the reflected light measurements. Contrast ratios $\leq 10^{-11}$ are needed to resolve rotational light curves of an Earth-like exoplanet \citep{palle2008identifying}, which should be achievable by a \emph{TPF}-type mission with high-contrast coronagraph or starshade \citep{ford2001characterization,trauger2007laboratory,turnbull2012search,cheng2015high}.

We will implicitly assume that planet $Q$ has both East-West and North-South albedo markings, and thus that the kernel geometry impacts the reflected light. In particular, we will assume two scenarios: perfect knowledge of the kernel, or kernel widths and changes in dominant colatitude that are constrained to $\pm 10^{\circ}$ and $\pm 20^{\circ}$, respectively, explained in Appendix \ref{sec:AlbVar}. Note that these uncertainties will depend on the planet's albedo contrast, and the photometric precision, in a non-linear way. We envision a triage approach for direct-imaging missions: planets that vary in brightness the most, and thus have the easiest albedo maps and spin axes to infer, will be the first for follow-up observations.

Of course, planetary radii are necessary to convert fluxes into apparent albedos \citep{qui2003earthshine,cowan2009alien}. Radii will likely be unknown, but could be approximated using mass-radius relations and mass estimates from astrometry or radial velocity, or inferred from bolometric flux using thermal infrared direct-imaging \citep[e.g. \emph{TPF-I};][]{beichman1999terrestrial,lawson2008terrestrial}. Real planets may also have variable albedo maps, e.g. short-term variations from changing clouds and smaller variations from long-term seasonal changes \citep{robinson2010detecting}, that could influence the apparent albedo on orbital timescales. These are difficulties that will be mitigated with each iteration of photometric detectors and theoretical models.

\subsection{Longitudinal Constraints}
\label{sec:long_con}
\begin{figure*}
	\centering
	\includegraphics[width=1.0\linewidth]{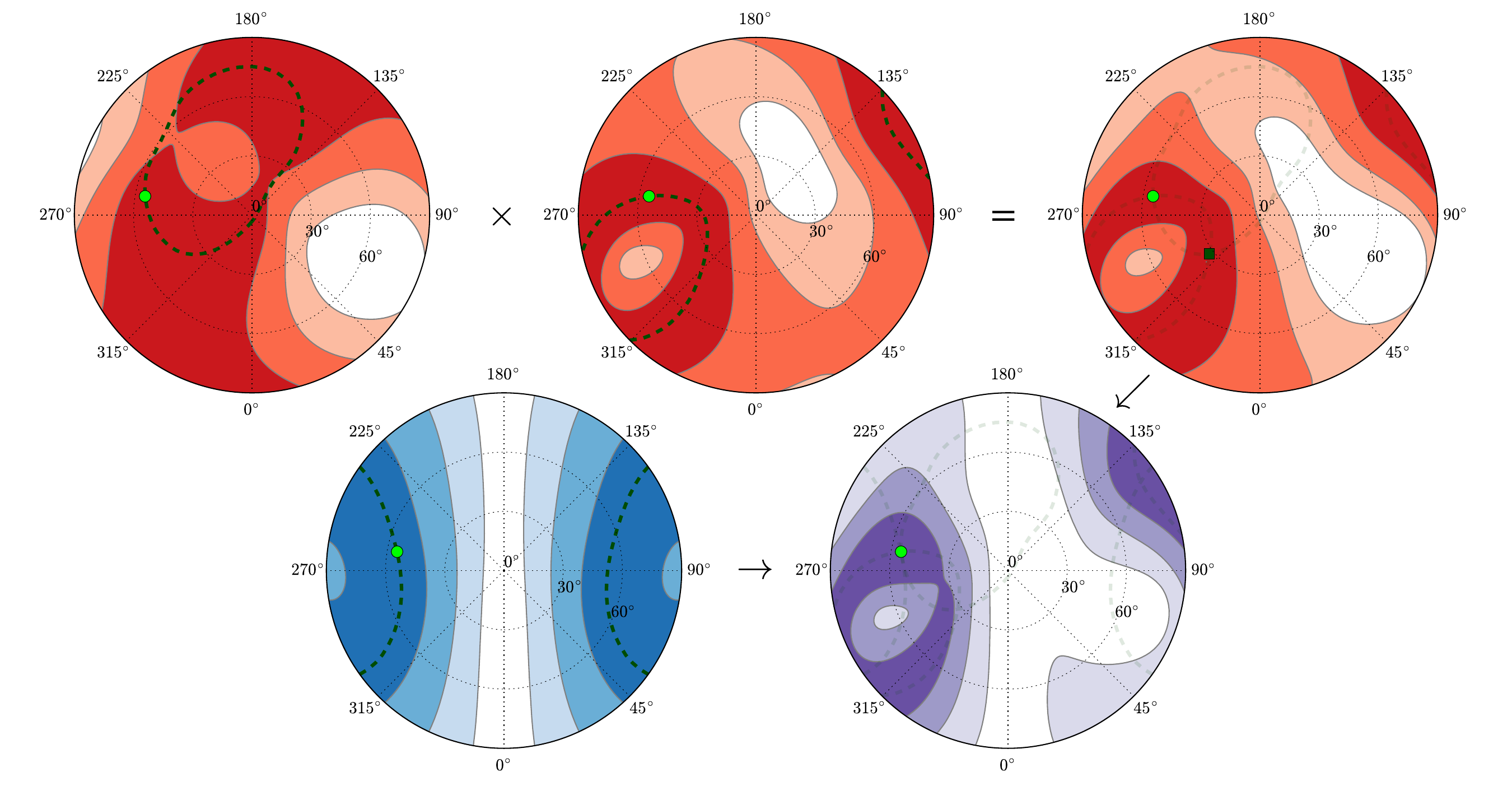}
	\caption{Predicted confidence regions for planet $Q$'s spin axis, from hypothetical single- and dual-epoch observations. The constraints are predicted using the kernel, as described in Sections \ref{sec:observing}--\ref{sec:joint_con} and Appendix \ref{sec:AlbVar}: longitudinal widths in the upper row (red), the change in dominant colatitude at the lower left (blue), and joint constraints at the lower right (purple). Obliquity is plotted radially: the center is $\Theta = 0^{\circ}$ and the edge is $\Theta = 90^{\circ}$. The azimuthal angle represents the planet's solstice phase. The green circles are the true planet $Q$ spin axis, while the dark dashed lines and square show idealized constraints assuming perfect knowledge of the orbital geometry and kernel (i.e. no uncertainties). The upper left and center panels describe planet $Q$ at $\xi(t) = \lbrace 120^{\circ},240^{\circ} \rbrace$, respectively, while the lower left panel incorporates both phases. For the colored regions, $10^{\circ}$ uncertainty is assumed on each kernel width, inclination, and orbital phase, while $20^{\circ}$ uncertainty is assumed on the change in dominant colatitude. Regions up to $3\sigma$ are shown, where darker bands are more likely. Observing a planet at just a few orbital phases can significantly constrain both its obliquity and axial orientation.}
	\label{fig:ex_obq_mega1020_1sigsPrfct}
\end{figure*}

A fit to the rotational light curve can be used to constrain the spin axis (and longitudinal map) of a planet with East-West albedo contrast. We can demonstrate these constraints using kernel widths in two ways, described in Section \ref{sec:observing} and shown in the upper panels of Figure \ref{fig:ex_obq_mega1020_1sigsPrfct}. The dark dashed lines and square are idealized constraints when assuming perfect knowledge of the orbital geometry and two kernel widths: $\sigma_{\phi 1} = 25.2^{\circ}$ at $\xi(t_{1}) = 120^{\circ}$, and \mbox{$\sigma_{\phi 2} = 51.7^{\circ}$ at $\xi(t_{2}) = 240^{\circ}$}. Alternatively, the red regions have  $10^{\circ}$ uncertainty on each width (Appendix \ref{sec:AlbVar}), where we use a normalized Gaussian probability density and include Gaussian weights for uncertainties on inclination and orbital phase.
	
The green circles represent the true planet $Q$ spin axis, which always lies on the idealized constraints. Only two spin axes are consistent with the ideal kernel widths from both orbital phases. We run more numerical experiments for a variety of system geometries (not shown) and find that perfect knowledge of the kernel width at three orbital phases uniquely determines the planetary spin axis. However, we find a degeneracy for planets with edge-on orbits, where two different spin configurations will produce the same kernel widths at all phases.

As anticipated, we also find planet $Q$'s spin axis (green circle) consistent with the uncertain kernel widths (dark red regions). Imperfect kernel widths at two phases allow all obliquities above $15^{\circ}$ at $1\sigma$, but exclude nearly one-fifth of spin axes at $3\sigma$. We find similar predictions for other orbital phases and planet parameters.

These examples also suggest that obliquity could be constrained for planets with variable albedo maps.  As long as albedo only changes on timescales longer than the rotational period, light curves will constrain both the instantaneous map and the planet's spin axis. A given spin orientation and orbital inclination dictates a specific kernel width as a function of orbital phase (left panel of Figure \ref{fig:ex_qtrace_wl260ph}), so we predict that light curves at three phases will be sufficient to pin down the planetary obliquity, even if the planet's map varies between phases.

\subsection{Latitudinal Constraints}
\label{sec:lat_con}
A fit to light curves from different orbital phases can be used to constrain the spin axis (and latitudinal map) of a planet with North-South albedo contrast. As described in Section \ref{sec:observing}, we can demonstrate this constraint using both perfect and uncertain knowledge of the change in dominant colatitude. Our predictions are shown in the lower left panel of Figure \ref{fig:ex_obq_mega1020_1sigsPrfct}. The idealized constraint here is $|\Delta\bar{\theta}_{12}| = 76.0^{\circ}$ between $\xi(t) = \lbrace 120^{\circ},240^{\circ} \rbrace$. Since the change in dominant colatitude is constrained between pairs of epochs, four orbital phases are needed to produce three independent constraints and uniquely determine planet $Q$'s spin axis. We find the same two-fold degeneracy as before for planets in edge-on orbits, even if one knows the change in dominant colatitude between all pairs of phases.

For the blue regions, we reapply our probability density from above and assume $20^{\circ}$ uncertainty on the change in dominant colatitude (Appendix \ref{sec:AlbVar}). The distribution is bimodal because only the \emph{magnitude} of the change can be constrained, not its direction. This means an observer would not know whether more Northern or Southern latitudes are probed at the later phase, affecting which spin axes are inferred.

\subsection{Joint Constraints}
\label{sec:joint_con}
For a planet with both East-West and North-South albedo contrast, one may combine longitudinal and latitudinal information to better constrain the planet's spin axis (and two-dimensional map). We show this for planet $Q$ in the lower right panel of Figure \ref{fig:ex_obq_mega1020_1sigsPrfct}. The idealized constraint shows that only the true spin configuration is allowed. We find this result for other system geometries---except using orbital phases $180^{\circ}$ apart, which creates a two-way degeneracy in the spin axis.

The confidence regions in purple assume our notional uncertainties on both kernel width and change in dominant colatitude (Appendix \ref{sec:AlbVar}). This prediction is not unimodal, but the $1\sigma$ region excludes obliquities below $30^{\circ}$. A distant observer would know that this planet's obliquity has probably not been eroded by tides \citep{heller2011tidal}, and that the planet likely experiences obliquity seasons.

\subsection{Pro/Retrograde Rotation}
\label{sec:pro_ret}
\begin{figure*}
	\centering
	\includegraphics[width=1.0\linewidth]{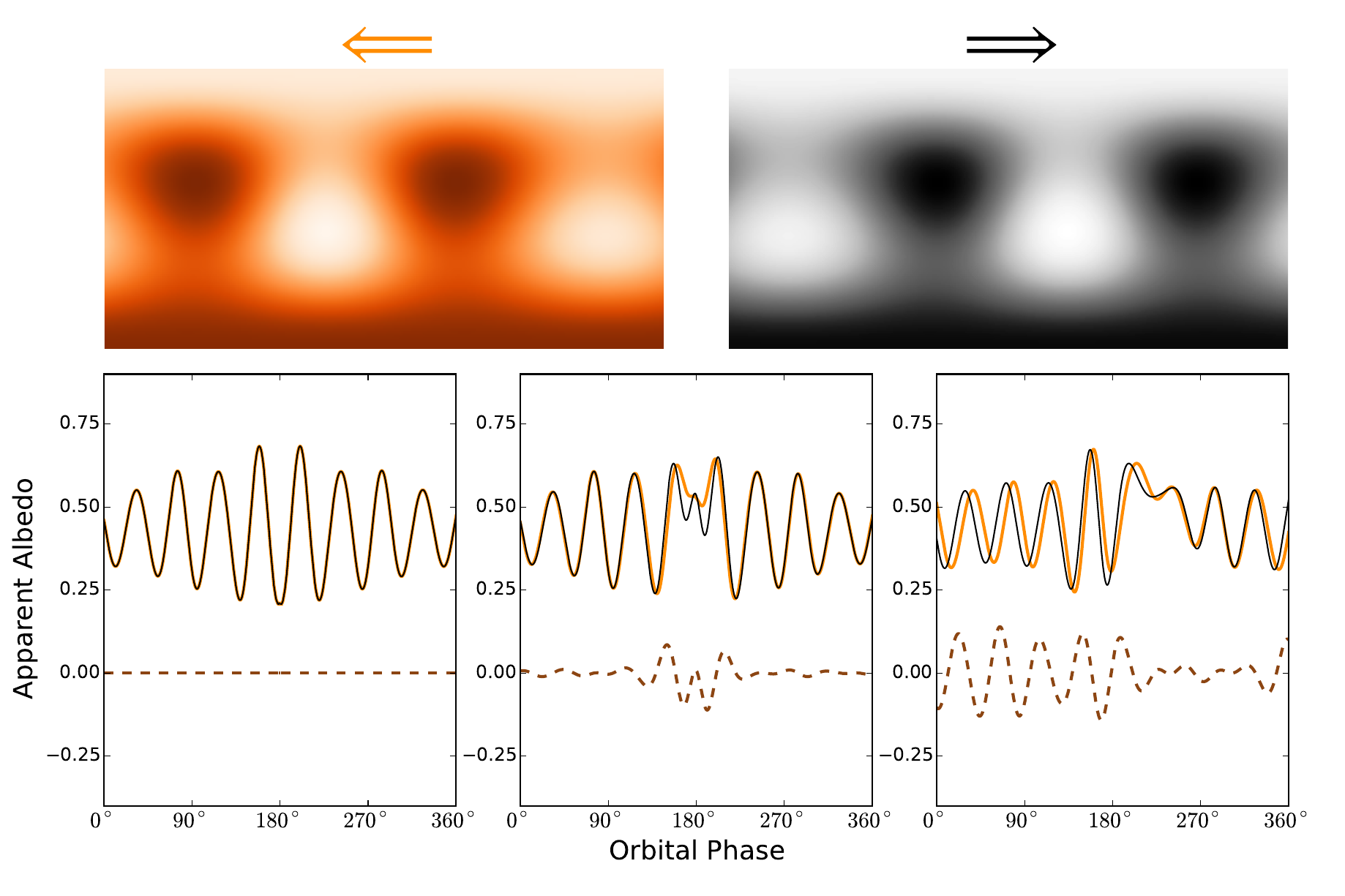}
	\caption{Apparent albedo as a function of orbital phase, shown for an edge-on, zero-obliquity planet on the left, planet $Q$ on the right, and an intermediate planet in the center. A low rotational frequency is used for clarity. The black and orange curves correspond to prograde and retrograde rotation, respectively, and the differences in apparent albedo are the dashed brown lines. Inferior conjunction occurs at $\xi(t) = 180^{\circ}$. The albedo maps are color-coded at the top, where arrows indicate spin direction and the prime meridians are centered. Note that these maps are East-West reflections of each other. The edge-on, zero-obliquity curves are identical, while the curves for the intermediate planet and planet $Q$ grow more distinct. Edge-on, zero-obliquity planets are hopeless, but one can distinguish pro/retrograde rotation for inclined, oblique planets by monitoring their brightness, particularly near crescent phases.}
	\label{fig:ex_AcurveB_edge_7525165_Q}
\end{figure*}

The sign of rotational angular frequency (positive = prograde) can affect the mean longitude of the kernel, but not its size and shape. There is a formal degeneracy for edge-on, zero-obliquity cases: prograde planets with East-oriented maps have identical light curves to retrograde planets with West-oriented maps. The motion of the kernel peak is the same over either version of the planet, implying the retrograde rotation in an inertial frame is \emph{slower} (\mbox{Appendix \ref{sec:peak_loc}}). We show this scenario in the left panel of Figure~\ref{fig:ex_AcurveB_edge_7525165_Q}, where the dashed brown line is the difference in prograde and retrograde apparent albedo. The orange and black planets are always equally bright because the same map features, in the upper panels, are seen at the same times.

However, the spin direction of oblique planets and/or those on inclined orbits may be deduced. Inclinations that are not edge-on most strongly alter a planet's light curve near inferior conjunction, seen in the center panel of Figure \ref{fig:ex_AcurveB_edge_7525165_Q}: this planet's properties are intermediate between the edge-on, zero-obliquity planet and \mbox{planet $Q$}. While a typical observatory's inner working angle would hide some of the signal, differences on the order of 0.1 in the apparent albedo would be detectable at extreme crescent phases. Alternatively, higher obliquity causes deviations that---depending on solstice phase---can arise around one or both quarter phases. This happens for planet $Q$ in the right panel of Figure \ref{fig:ex_AcurveB_edge_7525165_Q}, where both effects combine to distinguish the spin direction at most phases.
	
Inclination and obliquity influence apparent albedo because the longitudinal motion of the kernel peak is \emph{not} the same at all latitudes. One can break this spin degeneracy in principle, but we have not fully explored the pro/retrograde parameter space. In general, the less inclined and/or oblique a planet is, the more favorable crescent phases are for determining its spin direction.

\section{Conclusions}
\label{sec:conclude}
We have performed numerical experiments to study the problem of inferring a planet's obliquity from time-resolved photometry, for arbitrary albedo maps and viewing geometries. We have demonstrated that a planet's obliquity will influence its light curve in two distinct ways: one involving East-West albedo markings and another involving North-South markings. Provided this planet is not completely uniform, one could constrain both its albedo map and spin axis using reflected light.

The kernel---the product of visibility and illumination---has a peak, a longitudinal width, and a mean latitude that vary in time and are functions of viewing geometry. Analyzing the kernel enables us to predict constraints on a planet's spin axis from reflected light, including for maps that are East-West uniform (e.g. Jupiter-like) or North-South uniform (e.g. beach ball-like). Curiously, we find that kernel width offers better constraints on obliquity than dominant colatitude, suggesting that East-West albedo contrast is generally more useful than North-South contrast. This is partly because kernel width can be constrained even for variable albedo maps.

Furthermore, monitoring a planet at only a few epochs could determine its spin direction and significantly constrain its obliquity. In our case study of planet $Q$, we find crescent phases are favorable for telling prograde from retrograde rotation. Similarly, perfect knowledge of the kernel width at two orbital phases narrows the possible spin axes for planet $Q$ to two distinct configurations, while kernel width uncertainties of $10^{\circ}$ still exclude about three-quarters of spin axes at $1\sigma$. Adding the constraint on change in dominant colatitude between the same two phases completely specifies the true spin configuration of planet $Q$. A change in dominant colatitude with $20^{\circ}$ uncertainty excludes five-sixths of spin orientations at $1\sigma$.

Most importantly, we also find that perfect knowledge of the kernel width at just three phases, or its change in dominant colatitude between four phases, is generally sufficient to uniquely determine a planet's obliquity. This suggests that---in principle---rotational light curves at 2--4 distinct orbital phases uniquely constrain the spin axis of any planet with non-uniform albedo. This is good news for inferring the obliquity of planets with future direct-imaging missions.

\section*{Acknowledgments}
The authors thank the anonymous referee for important suggestions that improved the paper, and the International Space Science Institute (Bern, CH) for hosting their research workshop, ``The Exo-Cartography Inverse Problem." Participants included Ian M. Dobbs-Dixon (NYUAD), Ben Farr (U. Chicago), Will M. Farr (U. Birmingham), Yuka Fujii (TIT), Victoria Meadows (U. Washington), and Tyler D. Robinson (UC Santa Cruz). JCS was funded by an NSF GK-12 fellowship, and as a Graduate Research Trainee at McGill University.

\newpage

\footnotesize{
	\bibliographystyle{mn2e}
	\bibliography{JCS_Refs_Thesis}
}

\appendix

\section{Viewing Geometry}
\label{sec:v_geo}

\subsection{General Observer}
\label{sec:gen_obser}

The time-dependence of the kernel is contained in the sub-observer and sub-stellar angles: $\thobs$, $\phobs$ $\thst$, $\phst$. Since they do not depend on planetary latitude or longitude, these four angles may be factored out of the kernel integrals. However, the light curves are still functions of time, so we derive the relevant dependencies here.

\begin{figure}
	\centering
	\includegraphics[width=1.0\linewidth]{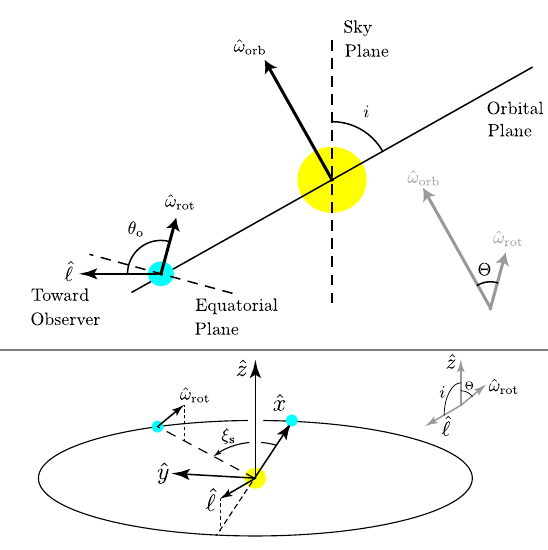}
	\caption{The upper panel shows a side view of the general planetary system. The rotational and orbital angular frequencies $\lbrace\wrot,\worb\rbrace$, \mbox{inclination $i$}, obliquity $\Theta$, sub-observer colatitude $\thobs$, and \mbox{observer viewing direction $\hat{\ell}$} are indicated. The lower panel is an isometric view, showing the solstice phase $\xis$. Superior conjunction occurs along the positive $x$-axis. Note the inertial coordinates, how they relate to the observer's viewpoint and planet's spin axis, and the angles between these vectors.}
	\label{fig:ObsStarDiag}
\end{figure}

In particular, we compute the sub-stellar and sub-observer locations for planets on circular orbits using seven parameters. Three are intrinsic to the system: rotational angular frequency, $\wrot \in (-\infty,\infty)$, orbital angular frequency, $\worb \in (0,\infty)$, and obliquity, $\Theta \in [0,\pi/2]$. Rotational frequency is measured in an inertial frame, where positive values are prograde and negative denotes retrograde rotation (for comparison, the rotational frequency of Earth is $\wrot^{\oplus}~\approx~2\pi/23.93\ \text{h}^{-1}$). Two more parameters are extrinsic and differ for each observer: orbital inclination, $i \in [0,\pi/2]$ where $i~=~90^{\circ}$ is edge-on, and solstice phase, $\xis \in [0,2\pi)$, which is the orbital angle between superior conjunction and the maximum Northern excursion of the sub-stellar point. The remaining parameters are extrinsic initial conditions: the starting orbital position, $\xi_{0} \in [0,2\pi)$, and the initial sub-observer longitude, $\phobsz \in [0,2\pi)$. These parameters are illustrated in Figure \ref{fig:ObsStarDiag}; other combinations are possible.

We define the orbital phase of the planet as $\xi(t) = \worb t + \xi_{0}$. Without loss of generality we may set the first initial condition as $\xi_{0}=0$, which puts the planet at superior conjunction when $t=0$. With no precession the sub-observer colatitude is constant, 
\begin{equation}
	\label{eq:tobs}
	\thobs(t) = \thobs. 
\end{equation}
This angle can be expressed in terms of the inclination, obliquity, and solstice phase using the spherical law of cosines (bottom of Figure \ref{fig:ObsStarDiag}):
\begin{equation}
	\label{eq:c_tob}
	\cos\thobs = \cos i\cos\Theta + \sin i\sin\Theta\cos\xis,
\end{equation}
\begin{equation}
	\label{eq:s_tob}
	\sin\thobs = \sqrt{1 - \cos^{2}\thobs}.
\end{equation}
The sub-observer longitude decreases linearly with time for prograde rotation, as we define longitude increasing to the East:
\begin{equation}
	\label{eq:phob}
	\phobs(t) = -\wrot t + \phobsz.
\end{equation}
The prime meridian ($\phi_{p} \Rightarrow \phi=0$) is a free parameter, which we define to run from the planet's North pole to the sub-observer point at $t=0$. This sets the second initial condition, namely $\phobsz=0$, and means
\begin{equation}
	\label{eq:c_phob}
	\cos\phobs = \cos(-\wrot t),
\end{equation}
\begin{equation}
	\label{eq:s_phob}
	\sin\phobs = \sqrt{1 - \cos^{2}\phobs}.
\end{equation}
Hence, the time evolution of the sub-observer point is specified by its colatitude and the rotational angular frequency.

The sub-stellar position is more complex for planets with non-zero obliquity. Consider an inertial Cartesian frame centered on the host star with fixed axes as follows: the $z$-axis is along the orbital angular \mbox{frequency, $\hat{z} = \hworb$,} while the $x$-axis points towards superior conjunction. The $y$-axis is then orthogonal to this plane using $\hat{y} = \hat{z} \times \hat{x}$ (bottom of \mbox{Figure \ref{fig:ObsStarDiag}).} In these inertial coordinates, the unit vector from the planet center towards the host star is $\rps = -\cos\xi\hat{x} - \sin\xi\hat{y}$. The corresponding unit vector from the star towards the observer is $\hat{\ell} = -\sin i\hat{x} + \cos{i}\hat{z}$. Our approach is to express everything in the inertial coordinate system, then find the sub-stellar point with appropriate dot products.

For the planetary surface, we use a second coordinate system fixed to the planet. We align the $z_{p}$-axis with the rotational angular frequency, ${\hat{z}_{p} = \hwrot}$, while the $x_{p}$-axis is set by our choice for the prime meridian (and initial sub-observer longitude.) The final axis, $y_{p}$, is again determined by taking $\hat{y}_{p} = \hat{z}_{p}~\times~\hat{x}_{p}$. We proceed in two steps, first finding the planetary axes when $t~=~0$, then using the planet's rotation to describe these axes at any time.

Since we disregard precession, the planet's rotation axis is time-independent:
\begin{equation}
	\hat{z}_{p} = \hwrot = -\cos\xis\sin\Theta\hat{x} - \sin\xis\sin\Theta\hat{y} + \cos\Theta\hat{z}. 
\end{equation}
The sub-observer point is on the prime meridian when $t=0$, so that
\begin{equation}
	\label{eq:yp0_def}
	\hat{y}_{p}(0) = \frac{\hat{z}_{p} \times \hat{\ell}}{\sin\thobs}.
\end{equation}
Computing this we find
\begin{equation}
	\label{eq:yp_0}
	\begin{aligned}
	\hat{y}_{p}(0) = \frac{1}{\sin\thobs} \Big[& -\cos i\sin\xis\sin\Theta\hat{x}\\
	& + (\cos i\cos\xis\sin\Theta -\sin i\cos\Theta)\hat{y}\\
	& -\sin i\sin\xis\sin\Theta\hat{z}\Big].
	\end{aligned}
\end{equation}
The starting $x_{p}$-axis is then found by taking $\hat{y}_{p}(0) \times \hat{z}_{p}$. The result is simplified by using Equation \ref{eq:c_tob}:
\begin{equation}
	\label{eq:xp_0}
	\begin{aligned}
	\hat{x}_{p}(0) = \frac{1}{\sin\thobs}\Big[& (\cos\xis\sin\Theta\cos\thobs - \sin i)\hat{x}\\
	& + \sin\xis\sin\Theta\cos\thobs\hat{y}\\
	& + \sin\Theta(\cos i\sin\Theta - \sin i\cos\xis\cos\Theta)\hat{z}\Big].
	\end{aligned}
\end{equation}

We can now find the planetary axes, in terms of the inertial axes, at any time by rotating Equations \ref{eq:yp_0} and \ref{eq:xp_0} about the $z_{p}$-axis:
\begin{equation}
	\hat{x}_{p} = \phantom{-}\cos(\wrot t)\hat{x}_{p}(0)+ \sin(\wrot t)\hat{y}_{p}(0),
\end{equation}
\begin{equation}
	\hat{y}_{p} = -\sin(\wrot t)\hat{x}_{p}(0)+ \cos(\wrot t)\hat{y}_{p}(0).
\end{equation}
The sub-stellar angles in the planetary coordinates may then be extracted from the relations
\begin{equation}
	\sin\thst\cos\phst = \rps\cdot\hat{x}_{p},
\end{equation}
\begin{equation}
	\sin\thst\sin\phst = \rps\cdot\hat{y}_{p},
\end{equation}
\begin{equation}
	\cos\thst = \rps\cdot\hat{z}_{p},
\end{equation}
resulting in
\begin{equation}
	\label{eq:c_thst}
	\cos\thst = \sin\Theta\cos\left[ \xi - \xis \right],
\end{equation}
\begin{equation}
	\label{eq:s_thst}
	\sin\thst = \sqrt{1 - \sin^{2}\Theta\cos^{2}\left[ \xi - \xis \right]},
\end{equation}
\begin{equation}
	\label{eq:c_phst}
	\cos\phst = \frac{\cos(\wrot t)a(t) + \sin(\wrot t)b(t)}{\sqrt{1 - \cos^{2}\thobs}\sqrt{1 - \sin^{2}\Theta\cos^{2}\left[ \xi - \xis \right]}},
\end{equation}
\begin{equation}
	\label{eq:s_phst}
	\sin\phst = \frac{-\sin(\wrot t)a(t) + \cos(\wrot t)b(t)}{\sqrt{1 - \cos^{2}\thobs}\sqrt{1 - \sin^{2}\Theta\cos^{2}\left[ \xi - \xis \right]}},
\end{equation}
where the factors $a(t)$ and $b(t)$ are given by
\begin{equation}
	a(t) = \Big\{\sin i\cos\xi - \cos\thobs\sin\Theta\cos\left[ \xi - \xis \right]\Big\},
\end{equation}
\begin{equation}
	b(t) = \Big\{\sin i\sin\xi\cos\Theta - \cos i\sin\Theta\sin \left[ \xi - \xis \right]\Big\}.
\end{equation}
Note that when $\thst = \left\lbrace 0,\pi \right\rbrace$, the sub-stellar longitude can be set arbitrarily to avoid dividing by zero in Equations \ref{eq:c_phst} and \ref{eq:s_phst}.

\begin{figure*}
	\centering
	\includegraphics[width=1.0\linewidth]{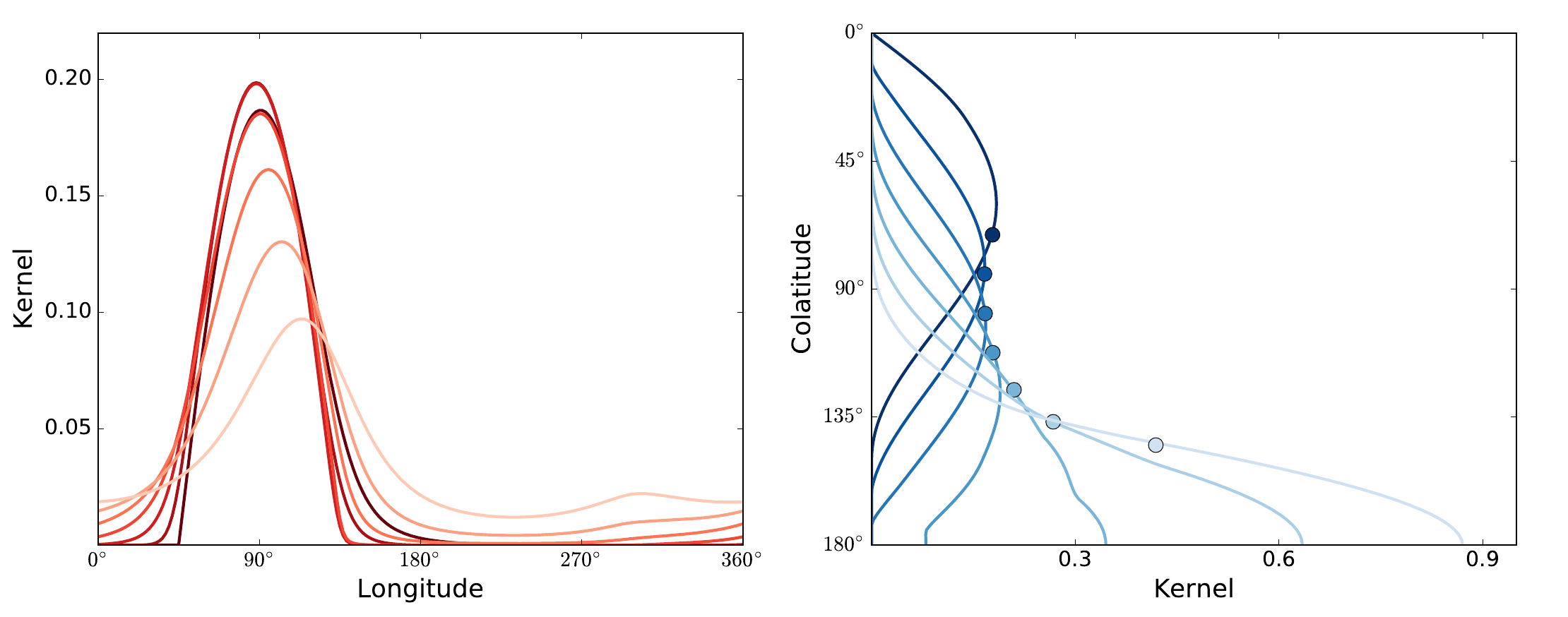}
	\caption{\emph{Left:} Longitudinal kernels of a planet with different obliquities at first quarter phase, $\xi(t) = 90^{\circ}$, where $\xis = 225^{\circ}$ and $i = 60^{\circ}$. The curves show $0^{\circ}$--$90^{\circ}$ obliquity in $15^{\circ}$ increments, where lighter shades of red indicate higher axial tilts. The kernel width of this planet changes as obliquity increases. \emph{Right:} Analogous latitudinal kernels that indicate the dominant colatitude, shown as a circle, also changes as obliquity increases.}
	\label{fig:ex_Kphith_225sol_obq}
\end{figure*}

\subsection{Polar Observer}
\label{sec:pole_obser}
Equations \ref{eq:c_phst} and \ref{eq:s_phst} for the sub-stellar longitude apply to most observers. However, the definition of $\hat{y}_{p}(0)$ in Equation \ref{eq:yp0_def} fails when the sub-observer point coincides with one of the planet's poles. Two alternate definitions can be used in these situations.

\emph{Case 1:} If the sub-stellar point \emph{will not} pass over the poles during orbit, we may define
\begin{equation}
	\hat{y}_{p}(0) = -\hat{y},
\end{equation}
so that
\begin{equation}
	\hat{x}_{p}(0) = \hat{y}_{p}(0)\times\hat{z}_{p} = -\cos\Theta\hat{x} - \cos\xis\sin\Theta\hat{z}.
\end{equation}
This results in
\begin{equation}
	\cos\phst = \frac{\cos(\wrot t)\cos\xi\cos\Theta + \sin(\wrot t)\sin\xi}{\sqrt{1 - \sin^{2}\Theta \cos^{2}\left[ \xi - \xis \right]}},
\end{equation}
\begin{equation}
	\sin\phst = \frac{-\sin(\wrot t)\cos\xi\cos\Theta + \cos(\wrot t)\sin\xi}{\sqrt{1 - \sin^{2}\Theta \cos^{2}\left[ \xi - \xis \right]}}.
\end{equation}

\emph{Case 2:} However, if the sub-stellar point \emph{will} pass over the poles during orbit, we define instead
\begin{equation}
	\hat{x}_{p}(0) = \hat{z},
\end{equation}
such that
\begin{equation}
	\hat{y}_{p}(0) = \hat{z}_{p}\times\hat{x}_{p}(0) = \cos\xis\hat{y}.
\end{equation}
This produces
\begin{equation}
	\cos\phst = \frac{-\sin(\wrot t)\sin\xi\cos\xis}{\sqrt{1 - \sin^{2}\Theta\cos^{2}\left[ \xi - \xis \right]}},
\end{equation}
\begin{equation}
	\sin\phst = \frac{-\cos(\wrot t)\sin\xi\cos\xis}{\sqrt{1 - \sin^{2}\Theta\cos^{2}\left[ \xi - \xis \right]}}.
\end{equation}

These special cases \emph{only} impact the sub-stellar longitude: expressions for the other angles are unchanged. As with a general observer, the Case 2 sub-stellar longitude may be set arbitrarily whenever $\thst = \left\lbrace 0,\pi \right\rbrace$.
 
\subsection{Zero Obliquity}
\label{sec:no_obq}
For non-oblique planets, $\Theta=0^{\circ}$, the sub-observer colatitude satisfies
\begin{equation}
	\cos\thobs = \cos i,
\end{equation}
\begin{equation}
	\sin\thobs = \sin i,
\end{equation}
while the sub-stellar angles become
\begin{equation}
	\cos\thst= 0,
\end{equation}
\begin{equation}
	\sin\thst = 1,
\end{equation}
\begin{equation}
	\cos\phst = \frac{\cos(\wrot t)c(t) + \sin(\wrot t)d(t)}{\sqrt{1 -\cos^{2}i}},
\end{equation}
\begin{equation}
	\sin\phst = \frac{-\sin(\wrot t)c(t) + \cos(\wrot t)d(t)}{\sqrt{1 - \cos^{2}i}},
\end{equation}
where $c(t)$ and $d(t)$ are given by
\begin{equation}
	c(t) = \sin i\cos\xi,
\end{equation}
\begin{equation}
	d(t) = \sin i\sin\xi.
\end{equation}
The sub-stellar longitude is therefore 
\begin{equation}
	\begin{aligned}
	\cos\phst &= \frac{\cos(\wrot t)\sin i\cos\xi + \sin(\wrot t)\sin i\sin\xi}{\sqrt{1 - \cos^{2}i}}\\
	&= \cos(\wrot t)\cos\xi + \sin(\wrot t)\sin\xi\\
	&= \cos(\xi - \wrot t),
	\end{aligned}
\end{equation}
\begin{equation}
	\begin{aligned}
	\sin\phst &= \frac{-\sin(\wrot t)\sin i\cos\xi + \cos(\wrot t)\sin i\sin\xi}{\sqrt{1 - \cos^{2}i}}\\
	&= -\sin(\wrot t)\cos\xi + \cos(\wrot t)\sin\xi\\
	&= \phantom{-}\sin(\xi - \wrot t).
	\end{aligned}
\end{equation}
In other words, $\thobs=i$, $\phobs=\phobsz-\wrot t$, $\thst=0$, and $\phst=\xi-\wrot t$.

\section{Kernel Details}
\label{sec:K_detail}

\subsection{Characteristics}
\label{sec:charact}

An important measure of the longitudinal kernel is its width, $\sigma_{\phi}$, as shown in the left panel of Figure \ref{fig:ex_Kphith_225sol_obq}. We treat this width mathematically as a standard deviation. Since $K(\phi,\mathbb{G})$ is on a periodic domain, we minimize the variance for each geometry with respect to the grid location of the prime meridian, $\phi_{p}$:
\begin{equation}
	\label{eq:std_width}
	\sigma^{2}_{\phi} = \text{min} \left[ \int_{0}^{2\pi} \left( \phi^{'} - \bar{\phi} \right)^{2} \hat{K}(\phi) \text{d}\phi \right]_{\phi_{p}},
\end{equation}
where $\hat{K}(\phi) = K(\phi) / \int K(\phi)\text{d}\phi$ is the spherically normalized longitudinal kernel, $\phi^{'}~\equiv~\phi + \phi_{p}$, and $\bar{\phi}$ is the mean longitude:
\begin{equation}
	\label{eq:mu_width}
	\bar{\phi} = \int_{0}^{2\pi} \phi^{'} \hat{K}(\phi) \text{d}\phi.
\end{equation}
All longitude arguments and separations in Equations \ref{eq:std_width} and \ref{eq:mu_width} wrap around the standard domain $[0,2\pi)$. Also note the \emph{unprimed} arguments inside the kernel: these make computing the variance simpler. The minimum variance determines the standard deviation of the kernel, and thus width, for a given geometry.

The dominant colatitude is similarly important for the latitudinal kernel, as shown in the right panel of Figure \ref{fig:ex_Kphith_225sol_obq}. \citet{cowan2012false} defined the dominant colatitude, $\bar{\theta}$, for a given geometry:
\begin{equation}
	\label{eq:Cow_domlat}
	\bar{\theta} = \oint \theta \hat{K}(\theta,\phi) \text{d}\Omega,
\end{equation}
where $\hat{K}(\theta,\phi) = K(\theta,\phi) / \oint K(\theta,\phi) \text{d}\Omega$ is the normalized kernel. \mbox{Equation \ref{eq:Cow_domlat}} is equivalent to
\begin{equation}
	\label{eq:dom_lat}
	\bar{\theta} = \int_{0}^{\pi} \theta \hat{K}(\theta) \sin\theta \text{d}\theta,
\end{equation}
where $\hat{K}(\theta) = K(\theta) / \int K(\theta)\sin\theta\text{d}\theta$ is the spherically normalized latitudinal kernel. The dominant colatitude is the North-South region that gets sampled most by the kernel (e.g. the circles in Figure \ref{fig:ex_Kphith_225sol_obq}.)

\begin{figure*}
	\centering
	\includegraphics[width=1.0\linewidth]{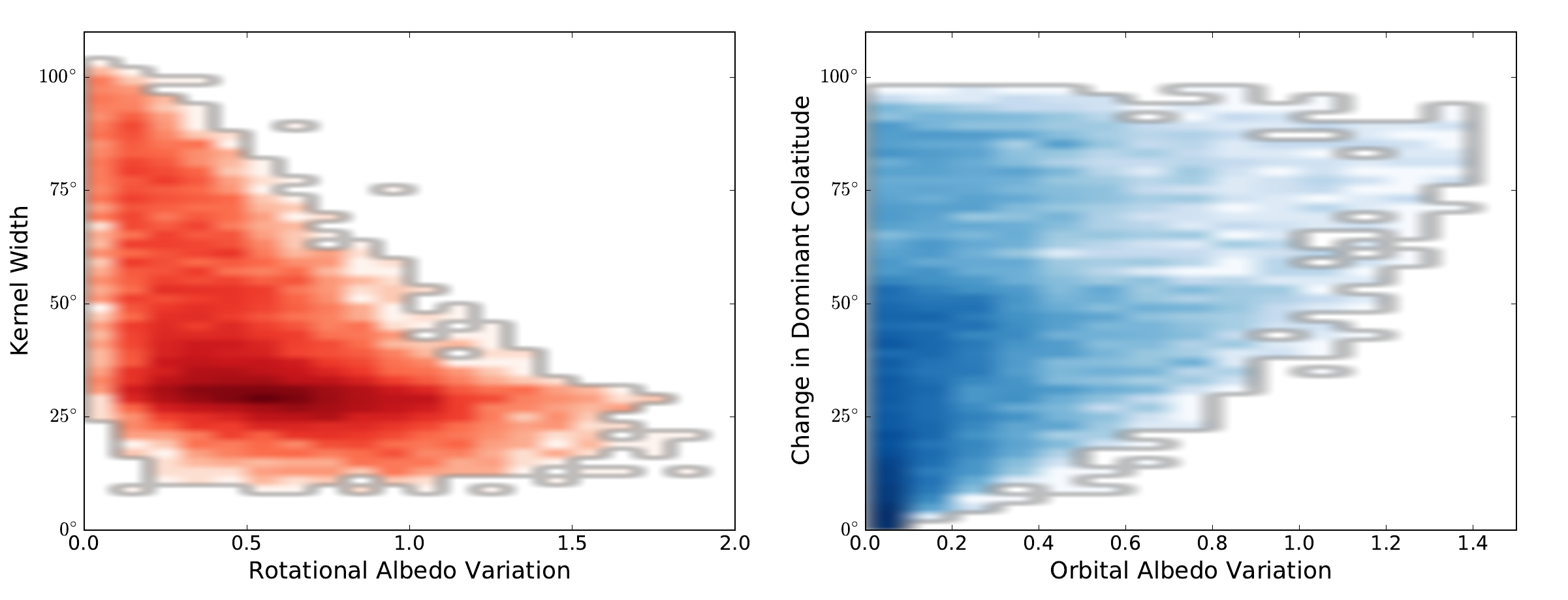}
	\caption{Kernel-albedo distributions for two sets of 10,000 planets generated via Monte Carlo, comparing rotational properties in red and orbital properties in blue. Changes in dominant colatitude are absolute values, and planets are binned in $0.1 \times 2^{\circ}$ regions in both panels. Each color scale is logarithmic: the darkest red and blue bins contain 230 and 279 planets, respectively. The scatter in either kernel characteristic decreases as the corresponding albedo variation rises. We can estimate the uncertainty on a kernel width or change in dominant colatitude by analyzing these distributions.}
	\label{fig:ex_MC_wdl_roVar}
\end{figure*}

\subsection{Albedo Variations}
\label{sec:AlbVar}

Figure \ref{fig:ex_LCs_edge_0q45q} demonstrates that obliquity can influence a planet's apparent albedo on both rotational and orbital timescales. Quantifying these relations helps predict the obliquity constraints we may expect from real observations. We use a Monte Carlo approach, simulating planets with different maps and viewing geometries. We generate albedo maps from spherical harmonics, $Y_{\ell}^{m}(\theta,\phi)$, on the same $101 \times 201$ grid in colatitude and longitude from Section \ref{sec:kern}:
\begin{equation}
	\label{eq:Ylm}
	A(\theta,\phi) = \sum_{\ell = 0}^{\ell_{\mathrm{max}}} \sum_{m = -\ell}^{\ell} C_{\ell}^{m} Y_{\ell}^{m}(\theta,\phi),
\end{equation}
where $\ell_{\mathrm{max}}$ is chosen to be 3, each coefficient $C_{\ell}^{m}$ is randomly drawn from the standard normal distribution, and the composite map is scaled to the Earth-like range $[0.1,0.8]$. Rotational and orbital changes in brightness are caused by East-West and North-South albedo markings, respectively, so we make three types of maps: East-West featured with $C_{\ell}^{m}(m \neq \ell) = 0$, North-South featured with $C_{\ell}^{m}(m \neq 0) = 0$, or no $C_{\ell}^{m}$ restrictions. For all maps with East-West features, we randomly offset the prime meridian. We generate 5,000 maps of each type.

For each map we randomly select an obliquity, solstice phase, inclination, and two orbital phases. Since inclination and orbital phase can be measured independent of photometry, we choose inclinations similar to \mbox{planet $Q$}, $i \in [50^{\circ}$,$70^{\circ}]$, and orbital phases $\lbrace \xi_{1},\xi_{2} \rbrace$ with \mbox{$\Delta\xi \in [110^{\circ},130^{\circ}]$}. Both phases are also at least $30^{\circ}$ from superior and inferior conjunction, which conservatively mimics an inner working angle at the selected inclinations. We assume the planet's rotational and orbital frequencies are known \citep{palle2008identifying,oakley2009construction}, and use the Earth-like \mbox{ratio $\wrot/\worb = 360$}. We divide roughly one planet rotation centered on each orbital phase into 51 time steps, then define the normalized amplitude of rotational and orbital albedo variations, $\Lambda_{\mathrm{rot}}$ and $\Lambda_{\mathrm{orb}}$, as
\begin{equation}
	\label{eq:ArotAmp}
	\Lambda_{\mathrm{rot}} = \frac{A_{\xi_{1}}^{\mathrm{high}} - A_{\xi_{1}}^{\mathrm{low}}}{\bar{A}_{\xi_{1}}},
\end{equation}
\begin{equation}
	\label{eq:AorbAmp}
	\Lambda_{\mathrm{orb}} = |\bar{A}_{\xi_{1}} - \bar{A}_{\xi_{2}}| \left(\frac{\bar{A}_{\xi_{1}} + \bar{A}_{\xi_{2}}}{2}\right)^{-1},
\end{equation}
where $A_{\xi_{1}}^{\mathrm{high}}$ and $A_{\xi_{1}}^{\mathrm{low}}$ are the extreme apparent albedos around the first phase, and $\bar{A}$ is the mean apparent albedo of all time steps around a given phase. For each computed $\Lambda_{\mathrm{rot}}$ and $\Lambda_{\mathrm{orb}}$, we calculate the corresponding kernel width and absolute value change in dominant colatitude, from Appendix \ref{sec:charact}. Figure \ref{fig:ex_MC_wdl_roVar} shows the resulting distributions, where rotational and orbital information is colored red and blue, respectively. We find similar results when relaxing constraints on the inclination and orbital phases.

We can estimate uncertainties on values of $\sigma_{\phi}$ and $|\Delta\bar{\theta}|$ using these distributions. The mean rotational and orbital variations are $\bar{\Lambda}_{\mathrm{rot}} \approx 0.54$ and $\bar{\Lambda}_{\mathrm{orb}} \approx 0.21$; the average kernel width and change in dominant colatitude are both roughly $38^{\circ}$. The full distributions have standard deviations of about $17^{\circ}$ in $\sigma_{\phi}$ and $24^{\circ}$ in $|\Delta\bar{\theta}|$, but roughly $5^{\circ}$ and $7^{\circ}$, respectively, when considering only large variations. To predict constraints on obliquity obtained from real data, we will assume there are single- and dual-epoch observations of planet $Q$ that have our mean variations $\bar{\Lambda}_{\mathrm{rot}}$ and $\bar{\Lambda}_{\mathrm{orb}}$. By considering samples only around these variations, we find about $10^{\circ}$ and $20^{\circ}$ standard deviations apiece in the kernel width and change in dominant colatitude. We use these standard deviations as uncertainties when creating the colored regions in Figure \ref{fig:ex_obq_mega1020_1sigsPrfct}.

\subsection{Peak Motion}
\label{sec:peak_loc}
Equations C1 and C2 from \citet{cowan2009alien} describe the motion of the kernel peak, where specular reflection occurs, for any planetary system. These equations can be written for edge-on, zero-obliquity planets using Section \ref{sec:no_obq}:
\begin{equation}
	\label{eq:cosspec}
	\cos\thspec = \frac{1 + \cos i}{\sqrt{2(1 + \cos i)}} = \frac{1}{\sqrt{2}},
\end{equation}
\begin{equation}
	\label{eq:tanspec}
	\begin{aligned}
	\tan\phispec &= \frac{\sin(\xi -\wrot t) + \sin(\phobsz - \wrot t)}{\cos(\xi - \wrot t) + \cos(\phobsz - \wrot t)}\\
	&= \frac{\sin(\worb t - \wrot t) + \sin(\phobsz - \wrot t)}{\cos(\worb t - \wrot t) + \cos(\phobsz - \wrot t)}.
	\end{aligned}
\end{equation}
When finding $\phispec$ from Equation \ref{eq:tanspec}, the two-argument arctangent must be used to ensure $\phispec \in [-\pi,\pi)$. This also means it is difficult to simplify the equation with trigonometric identities.

Instead, we can explicitly write the argument of Equation \ref{eq:tanspec} in terms of the first meridian crossing, $\xi_{m}$, the earliest orbital phase after superior conjunction that the kernel peak recrosses the prime meridian:
\begin{equation}
	\label{eq:argspec}
	\phi_{\text{spec}}(\xi;\xi_{m}) = \mp\frac{\pi}{2}\left(4\frac{\xi}{\xi_{m}} + \left[ 1-\text{sgn}\left(\cos\frac{\xi}{2}\right)\right]\right),
\end{equation}
where the leading upper sign applies to prograde rotation and vice versa. The first meridian crossing is related to the planet's frequency (or period) ratio by
\begin{equation}
	\label{eq:wperxione}
	\left|\frac{\wrot}{\worb}\right| = \left|\frac{P_{\text{orb}}}{P_{\text{rot}}}\right| = \frac{1}{2}\left( \frac{4\pi}{\xi_{m}} \pm 1 \right),
\end{equation}
while the number of solar days per orbit is
\begin{equation}
	\label{eq:nsol}
	N_{\text{solar}} = \left|\frac{\wrot}{\worb}\right| \mp 1,
\end{equation}
following the same sign convention. Note that the frequency/period ratios and the number of solar days do not have to be integers. We reiterate that Equations \ref{eq:cosspec}--\ref{eq:nsol} apply to \emph{edge-on, zero-obliquity} planets.

Equation \ref{eq:wperxione} gives two frequency ratios for each first meridian crossing, one prograde and another retrograde that is smaller in magnitude by unity. Equation \ref{eq:nsol} then states the corresponding difference in solar days is unity but reversed, making the longitudes of both kernel peaks in \mbox{Equation \ref{eq:argspec}} analogous at each orbital phase. These two versions of the planet have East-West mirrored albedo maps and identical light curves: they are formally degenerate. An inclined, oblique planet has pro/retrograde versions that could be distinguished, as discussed in Section \ref{sec:pro_ret}.

\bsp

\label{lastpage}

\end{document}